\documentclass[prd,reprint]{revtex4-1}
\usepackage{graphicx}
\usepackage{array} % this is needed for automatic commands applied to table columns
\def\sect#1{Section~\ref{#1}}

\def\>{\rangle}
\def\<{\langle}
\def\etal{\textit{et~al}}

\newcommand{\rf}[1]{(\ref{#1})}

\def\etab{\end{tabular}}  
  
\def\bit{\begin{itemize}}
\def\eit{\end{itemize}}
\def\bml{\begin{multline}}
\def\eml{\end{multline}}
\def\be{\begin{equation}} 
\def\ee{\end{equation}}  
\def\bea{\begin{eqnarray}}  
\def\eea{\end{eqnarray}} 
\def\bes{\begin{equation*}} 
\def\ees{\end{equation*}}  
\def\beas{\begin{eqnarray*}}  
\def\eeas{\end{eqnarray*}} 
\def\bmu{\begin{multline}}
\def\emu{\end{multline}}
\def\bal{\begin{array}{l}} 	%for equations which are at least three lines long, 
\def\eal{\end{array}}

\def\MeV{\textrm{ MeV}}

\newcommand{\ve}[1]{\mathbf{#1}} 	%roman vector
\def\letterS{S}
\def\letterP{P}
\def\letterD{D}
\def\letterF{F}
\def\greek#1{\def\letter{#1}
\ifx\letter\letterS\Sigma
\else
	\ifx\letter\letterP\Pi
	\else
		\ifx\letter\letterD\Delta
		\else
			\ifx\letter\letterF\Phi
			\else XXX
			\fi
		\fi
	\fi
\fi
}

\def\qq{q\bar q}
\def\QQ{Q\bar Q}
\def\qQ{q\bar Q}
\def\Qq{Q\bar q}

\newcommand{\ket}[1]{|           {#1}           \rangle}
\newcommand{\bra}[1]{\langle           {#1}           |}

\renewcommand{\u}[1]{\rm{#1}}
\def\cn#1#2#3{^{#1}\u{#2}_{#3}}     %normal chemical notation
\def\an{\cn} %atomic notation
\newcommand{\uS}{\rm{S}}
\newcommand{\uP}{\rm{P}}

\def\epem{e^+e^-}
\def\xQN#1,{\def\Element{#1}%
\ifx\Element\endpiece
\else 
	\if\Element/\def\ApplyOrLiteral{1} % I could use ; : /
	\else
		\if\ApplyOrLiteral0\ElementRule\Element %Note, this used to be \ElementRule{\Element} !!!! thus can't have commands within the elementrule, but can have defs
		\else\Element\def\ApplyOrLiteral{0}
		\fi
	\fi
	\expandafter\xQN
\fi}

\def\xNQ#1,{\def\ElementRule{#1}%
\ifx\ElementRule\endpiece
\else 
	\if\ElementRule/\def\ApplyOrLiteral{1} % I could use ; : /
	\else
		\if\ApplyOrLiteral0\ElementRule\Element %Note, this used to be \ElementRule{\Element} !!!!
		\else\ElementRule\def\ApplyOrLiteral{0}
		\fi
	\fi
	\expandafter\xNQ
\fi}

\def\k|#1>{\ket{#1}} % there is no real need to define anything fancy for kets
\def\b<#1|{\bra{#1}} % or bras because i can just use \k|\QN{}{}>

\def\bk<#1|#2>{\langle #1 | #2 \rangle}
\def\Bk#1<#2|#3>{\def\ElementRule{#1}\def\ApplyOrLiteral{0}
\xQN/,\langle,#2,xxx,\vert #3 \rangle}
\def\bK#1<#2|#3>{Undefined}

\def\BK#1<#2|#3>{\def\ElementRule{#1}\def\ApplyOrLiteral{0}
\xQN/,\langle,#2,/,|,#3,/,\rangle,xxx,}
\def\KB<#1|#2>#3{\def\Element{#3}\def\ApplyOrLiteral{0}
\xNQ/,\langle,#1,/,|,#2,/,\rangle,xxx,}

\def\me<#1|#2|#3>{\langle #1\vert #2\vert #3\rangle}
\def\ME#1<#2|#3|#4>{\def\ElementRule{#1}\def\ApplyOrLiteral{0}
\xQN/,\langle,#2,/,\vert,/,#3,/,\vert,#4,/,\rangle,xxx,}
\def\EM<#1|#2|#3>#4{\def\Element{#4}\def\ApplyOrLiteral{0}
\xNQ/,\langle,#1,/,\vert,/,#2,/,\vert,#3,/,\rangle,xxx,}

\def\rme<#1|#2|#3>{\langle #1\Vert #2\Vert #3\rangle}
\def\RME#1<#2|#3|#4>{\def\ElementRule{#1}\def\ApplyOrLiteral{0}
\xQN/,\langle,#2,/,\Vert,/,#3,/,\Vert,#4,/,\rangle,xxx,}
\def\EMR<#1|#2|#3>#4{\def\Element{#4}\def\ApplyOrLiteral{0}
\xNQ/,\langle,#1,/,\Vert,/,#2,/,\Vert,#3,/,\rangle,xxx,}

\def\endpiece{xxx}

\def\epemwidth#1{\Gamma_{{\epem}\to{#1^3 \rm S_1}}}
\def\etab#1{\eta_b(#1\uS)}
\def\hyperfine#1{\Delta M_{#1\uS}}

\def\hs{hyperfine splitting}

\begin{document}

\title{Hyperfine splitting and the experimental candidates for $\etab{2}$}


\author{T.~J.~Burns }
\email{{\tt t.burns@oxon.org}\href{mailto:t.burns@oxon.org}}
\affiliation{Department of Mathematical Sciences, Durham University, Durham DH1 3LE, United Kingdom}
\begin{abstract}

Predictions for the hyperfine splitting of 2S bottomonia are compared with the two recent experimental candidates for the $\etab{2}$. The smaller splitting of the Belle state is consistent with unquenched lattice QCD computations, many potential models, and a model-independent relation which works well for charmonia. The larger splitting for the state extracted from CLEO data is inconsistent with most predictions. 

\pacs{14.40.Pq,12.38.Gc,12.39.Jh,12.39.Pn}
%14.40.Pq 	Heavy quarkonia
%12.38.Gc 	Lattice QCD calculations (see also 11.15.Ha Lattice gauge theory)
%12.39.Jh 	Nonrelativistic quark model
%12.39.Pn 	Potential models

\end{abstract}


\maketitle

%\thispagestyle{fancy}

%\tableofcontents

\section{Introduction}

Within a short space of time, two groups have independently claimed the discovery of the $\eta_b(2\uS)$. In an analysis of CLEO data, Dobbs \etal.\ \cite{Dobbs:2012zn} observe an enhancement in the M1 decay $\Upsilon(2\uS)\to \etab{2}\gamma$, with mass and hyperfine splitting $\Delta M_{2\uS}=M_{\Upsilon{(2\uS)}}-M_{\eta_b{(2\uS)}}$:
\bea
M_{\etab{2}}&=& 9974.6 \pm 2.3\pm 2.1\MeV,\\
\hyperfine{2}&=&48.7 \pm 2.3 \pm 2.1 \MeV.
\eea
The Belle collaboration \cite{Mizuk:2012pb} reports a state in the E1  decay $h_b(2\uP)\to \etab{2}\gamma$, with mass and hyperfine splitting:
\bea
M_{\etab{2}}&=& 9999.0\pm 3.5 ^{+2.8}_{-1.9}\MeV,\\
\hyperfine{2}&=&24.3^{+4.0}_{-4.5} \MeV.
\eea
The striking disagreement in the masses has not yet been addressed in the literature. The aim of this paper is to compare the conflicting results with predictions from a range of theoretical approaches:  lattice QCD (\sect{lattice}), a model-independent mass relation (\sect{relation}), potential models (\sect{potential}), and the unquenched quark model (\sect{unquenched}). %Experimental channels in which the $\etab{2}$ may be confirmed are discussed (\sect{confirming}).

The data which forms the basis of the discussion is collected in Table \ref{tabl}, a compilation of experimental values and theoretical predictions for both the 1S and 2S hyperfine splittings of bottomonia. It is useful to look at the 1S and 2S splittings together, because the two are not independent: several of the lattice QCD predictions for the 2S splitting, and the prediction of the model-independent mass relation, are normalised against the 1S splitting, and in potential models the 1S and 2S splittings are controlled by the same parameters of the spin-spin interaction. 
%A clear correlation between the 1S and 2S splittings is evident in Figure \ref{fig}, a plot of the data in Table \ref{tabl}.

The role of the 1S splitting in the interpretation of predicted 2S splittings is particularly important because its experimental value is still in flux. Various measurements and averages for the 1S splitting are shown in Table \ref{tabl}. The 2012 PDG value (``PDG12'') is obtained from an average of three experimental results in  $\Upsilon(3\uS)\to \etab{1}\gamma$ \cite{Aubert:2008ba,Bonvicini:2009hs} and $\Upsilon(2\uS)\to \etab{1}\gamma$~\cite{Aubert:2009as}. More recently Belle~\cite{Collaboration:2011ch} observed $h_b(1\uP)\to\eta_b(1\uS)\gamma$  and their $\etab{1}$ mass corresponds to a hyperfine splitting with central value 10.0~MeV lower than PDG12 . In a mini-review in the 2012 PDG \cite{pdg12,Eidelman:2012vu}, a new average value for the 1S splitting (``PDG12*'' in Table \ref{tabl}) is computed taking account of the Belle measurement, almost 5 MeV smaller than the previous average.

Dobbs \etal.\ and Belle also observe the $\etab{1}$ in the analogous radiative decays in which they claim the $\etab{2}$; their measured values for the 1S \hs~ are also shown in Table \ref{tabl}. In an analysis of $\Upsilon(1\uS)\to \eta_b(1\uS)\gamma$ Dobbs \etal.\ obtain a 1S splitting which is consistent with PDG12 and PDG12*. In $h_b(1\uP,2\uP)\to \eta_b(1\uS)\gamma$ Belle obtains a 1S splitting which is smaller than both PDG12 and PDG12*, but consistent with the earlier Belle \cite{Collaboration:2011ch} measurement.

The important point is that each of the experimental measurements which post-date the computation of PDG12 have a smaller central value for the 1S splitting. This is particularly relevant to the analysis of the lattice QCD predictions, which is the first topic of discussion.

\begin{table}
\begin{ruledtabular}
 \begin{tabular}{l >{$}l<{$}>{$}l<{$}}
				 			&\hyperfine{1}  		& \hyperfine{2}		\\
\hline%\multicolumn{2}{l}{\textit{Experiment}}				&				\\
\textit{Experiment}					&				&				\\
PDG12 				\cite{pdg12}		&69.3\pm 2.8			&				\\		
Belle \cite{Collaboration:2011ch}			&59.3\pm 1.9^{+2.4}_{-1.4}	&				\\
PDG12* 			\cite{pdg12,Eidelman:2012vu}	&64.5\pm 3.0			&				\\		
Dobbs \etal.\ 				 \cite{Dobbs:2012zn}	&67.1\pm	4.1		&	48.7\pm2.3\pm2.1		\\
Belle				 \cite{Mizuk:2012pb}	&57.9\pm	2.3		&	24.3^{+4.0}_{-4.5}	\\
\hline
\textit{Lattice QCD}		 			&				&				\\
Gray \etal.\		 	  \cite{Gray:2005ur} 	& 61\pm 14 			& 30\pm 19 \\ 
Meinel		\cite{Meinel:2010pv}	 		&  60.3\pm7.7\dag 		& 23.5\pm4.7\dag \\ 
Meinel 		\cite{Meinel:2010pv}			& 		 		& 28.0\pm4.2 \ddag \\
%Meinel		\cite{Meinel:2010pv}	 		& (57.9\pm2.3)	 		& 23.3\pm4.3???  \\ 
Dowdall \etal.\ 	\cite{Dowdall:2011wh}			& 70 \pm 9 			& 35 \pm 3\ddag 	 \\
%Dowdall \etal.\ 	\cite{Dowdall:2011wh}			& (57.9\pm2.3) 			& 29 \pm 4 	 \\
Lewis and Woloshyn   \cite{Lewis:2011ti,Lewis:2012ir} 	&  56\pm 1			& 24 \pm 3 \\ 
Burch and Ehmann	\cite{Burch:2007fj}		&37\pm	11			&13\pm	26\\
Burch and Ehmann	\cite{Burch:2007fj}		&71\pm	8\diamond			&27\pm	17\diamond\\
\hline
\textit{Potential models and related}			&				&	\\
%BJ(82)  ? & 27 & 13 \\ 
%CGM(78)  ? & 90 & 40 \\ 
Badalian \etal.\ \cite{Badalian:2009cx} & 64.2 \pm 0.4& 36 \pm 2\\ 
Badalian \etal.\ \cite{Badalian:2009cx} & 70.0\pm 0.4& 36 \pm 2\\ 
Badalian \etal.\ \cite{Badalian:2010uj} & 63.4 & 36 \pm 2\\ 
Badalian \etal.\ \cite{Badalian:2010uj} & 71.1 & 36 \pm 2\\ 
Bali \etal.\ \cite{Bali:1997am} & 44 & 27 \\ 
Bali \etal.\ \cite{Bali:1997am} & 50 & 30 \\ 
Bali \etal.\ \cite{Bali:1997am} & 89 & 47 \\ 
%Bander \etal.\ \cite{Bander:1983ew} & 49 & 18 \\ 
Bander \etal.\ \cite{Bander:1983ew} & 57 & 21 \\ 
Bander \etal.\ \cite{Bander:1983ew,Bander:1984vr} & 60 & 22 \\ 
Bander \etal.\ \cite{Bander:1983ew} & 58 & 21 \\  
Buchmuller and Tye \cite{Buchmuller:1980su} & 46 & 23 \\ 
Chen and Oakes \cite{Chen:2000ej} & 97.7 & 39.6 \\ 
%Chen and Oakes \cite{Chen:2000ej} & 38.2 & 19.8 \\ 
Chen and Oakes \cite{Chen:2000ej} & 47.6 & 22.4 \\ 
Chen and Oakes \cite{Chen:2000ej} & 54.3 & 25.6 \\ 
Ebert \etal.\ \cite{Ebert:2002pp} & 60 & 30 \\ 
Eichten and Feinberg \cite{Eichten:1980mw} & 94.9 & 41.2 \\ 
Eichten and Quigg \cite{Eichten:1994gt} & 87 & 44 \\ 
Fulcher \cite{Fulcher:1988ca} & 43 & 26 \\ 
Fulcher \cite{Fulcher:1990kx} & 92 & 44 \\ 
Fulcher \cite{Fulcher:1991dm} & 46 & 23 \\ 
Giachetti and Sorace	\cite{Giachetti:2012tw}	&75.71&	37.90\\
Godfrey and Isgur \cite{Godfrey:1985xj} & 63 & 27 \\ 
Grotch \etal.\ \cite{Grotch:1984gf} & 67 & 31 \\ 
%Grotch \etal.\ \cite{Grotch:1984gf} & 78 & 37 \\ 
Gupta \etal.\ \cite{Gupta:1982kp} & 35 & 19 \\ 
Gupta \etal.\ \cite{Gupta:1986xt} & 44 & 26 \\ 
Gupta \etal.\ \cite{Gupta:1989jd} & 47.8 & 23.3 \\ 
Gupta and Johnson \cite{Gupta:1995ps} & 52.7 & 25.6 \\ 
Ito \cite{Ito:1990he} & 63 & 23 \\ 
Lahde \etal.\ \cite{Lahde:1998ee} & 79 & 44 \\ 
Li and Chao \cite{Li:2009nr} & 71 & 29 \\ 
McClary and Byers \cite{McClary:1983xw} & 101 & 40 \\ 
Motyka and Zalewski \cite{Motyka:1997di} & 56.7 & 28 \\ 
Moxhay and Rosner\cite{Moxhay:1983vu,Rosner:1983jc} & 57 & 26 \\ 
Ng \etal.\ \cite{Pantaleone:1985uf} & 35 & 19 \\ 
Ono and Schoberl \cite{Ono:1982ft} & 80 & 31 \\ 
%Pantaleone \etal.\ \cite{Pantaleone:1985uf} & 41 & 22 \\ 
Pantaleone \etal.\ \cite{Pantaleone:1985uf} & 46 \pm 3& 23\pm 1 \\ 
%Pantaleone \etal.\ \cite{Pantaleone:1985uf} & 30 & 18 \\ 
%Pantaleone \etal.\ \cite{Pantaleone:1985uf} & 35 & 19 \\ 
Patel and Vindokumar \cite{Patel:2008na} & 58 & 33 \\ 
Patel and Vindokumar \cite{Patel:2008na} & 60 & 38 \\ 
%Radford and Repko \cite{Radford:2007vd} & 39.26 & 19.9 \\ 
Radford and Repko \cite{Radford:2007vd} & 46.99 & 23.81 \\ 
Radford and Repko \cite{Radford:2009qi} & 67.5\pm 0.7 & 35.9\pm 0.3\\ 
Recksiegel and Sumino \cite{Recksiegel:2003fm} & 45 \pm 11& 22\pm 8 \\ 
Shah \etal.\ \cite{Shah:2012js} & 68.00 & 32.45 \\ 
Zeng \etal.\ \cite{Zeng:1994vj} & 48.9 & 25.6 \\ 
Zhang \etal.\ \cite{Zhang:1991et} & 48 & 28 \\ 
 \end{tabular}
\caption{Experimental data and theoretical predictions for the 1S and 2S hyperfine splittings (in MeV) of bottomonia.\\
$\dag$ Normalised to the experimental 1P tensor splitting.\\
$\ddag$ Normalised to the PDG12 1S splitting.\\
$\diamond$ Quenched lattice QCD.}
\label{tabl}
\end{ruledtabular}
\end{table}

\section{Lattice QCD}\label{lattice}

Various predictions for the 1S and 2S splittings in nonrelativistic lattice QCD are collected in Table \ref{tabl}. Apart from those marked ($\diamond$), all of the results are from unquenched lattice QCD. 

For the 1S splitting, Gray \etal.\ \cite{Gray:2005ur} (pre-dating the discovery of the $\etab{1}$) and  Meinel  \cite{Meinel:2010pv} are consistent with all of the measured values and the PDG averages; Dowdall \etal.\ \cite{Dowdall:2011wh} and the quenched result of Burch and Ehmann \cite{Burch:2007fj} are consistent with the PDG averages and Dobbs \etal., while Lewis and Woloshyn \cite{Lewis:2011ti,Lewis:2012ir} are consistent with Belle. 

The predictions for the 2S splitting discriminate more strongly. The result of Gray \etal.\ \cite{Gray:2005ur}, with its somewhat larger errors, is consistent with both Dobbs \etal.\ and Belle. Otherwise, none of the predictions are consistent with Dobbs \etal. The predictions of Meinel \cite{Meinel:2010pv} (one normalised to the 1S splitting, the other to the 1P splitting), Lewis and Woloshyn \cite{Lewis:2011ti,Lewis:2012ir}, and Burch and Ehmann \cite{Burch:2007fj} are all consistent with Belle. The prediction of Dowdall \etal.\ \cite{Dowdall:2011wh} falls between the measured values of Belle and Dobbs \etal.\ and is not consistent with either. 

Two of the predictions for the 2S splitting, marked in the table ($\ddag$), are normalised to the PDG12 value of the 1S splitting. As discussed previously, recent experiments \cite{Collaboration:2011ch,Mizuk:2012pb,Dobbs:2012zn} all measure a smaller value for the 1S splitting, which would imply that these lattice predictions are overestimates. This is particularly interesting in the case of Dowdall \etal.\ \cite{Dowdall:2011wh}, the only lattice prediction which is inconsistent with Belle.

The quantity measured on the lattice is the ratio of 2S to 1S splittings. Dowdall \etal.\ obtain
\be
\frac{\hyperfine{2}}{\hyperfine{1}}=0.499\pm0.042,\label{lattice[1]}
\ee
while Meinel obtains (with errors added in quadrature),
\be
\frac{\hyperfine{2}}{\hyperfine{1}}=0.403\pm0.059.\label{lattice[2]}
\ee
 
By normalising against the updated PDG12* 1S splitting, rather than PDG12, the prediction of Dowdall \etal.\ is brought into agreement with Belle, $\hyperfine{2}=32.2\pm 4.2$~MeV, and the agreement between Meinel and Belle is improved, $\hyperfine{2}=25.9\pm5.0$~MeV.

An alternative interpretation of the results of Dowdall \etal.\ and Meinel is to normalise against their computed values for the 1S splitting. The resulting predictions are consistent with Belle,  $\hyperfine{2}=34.9\pm 7.4$~MeV and $\hyperfine{2}=24.3\pm 6.7$~MeV respectively.

It is also interesting to normalise Dowdall \etal.\ and Meinel against the most accurate measurement of the 1S splitting from a single experiment, that of Belle \cite{Mizuk:2012pb}. The predicted 2S splittings are consistent with the measured value at Belle despite smaller errors, $\hyperfine{2}=28.9\pm 3.6$~MeV and $\hyperfine{2}=23.3\pm 4.3$~MeV respectively.

(Notice that the adjustment to the Meinel value brings it into better agreement with the other value of Meinel quoted in Table \ref{tabl}, which is normalised to the 1P tensor splitting. This implies that the ratio of 1S hyperfine to 1P tensor splittings, which Meinel also computes directly on the lattice, is in better agreement with experiment using the 1S of Belle than with PDG12, which is interesting in its own right. Intriguingly, the computed masses of $B_c$ and $B_s$ mesons in Ref. \cite{McNeile:2012qf} are in better agreement with experiment when normalised to the Belle $\etab{1}$ mass rather than that of the PDG, although  it is not a statistically significant effect.)

A more direct interpretation of the above lattice results is to compare the computed ratio of splittings directly with experiment, rather than extracting a prediction for the 2S splitting which depends upon (and is subject to the error in) the 1S splitting against which it is normalised. In this context, two other lattice results are worth mentioning. Gray \etal.\ \cite{Gray:2005ur} measure the ratio
\be
\frac{\hyperfine{2}}{\hyperfine{1}}=0.5\pm 0.3\label{lattice[3]},
\ee
while the quenched computation of Davies \etal.\ \cite{Davies:1998im}  has the ratio
\be
\frac{\hyperfine{2}}{\hyperfine{1}}=0.49\pm0.09\label{lattice[4]}.
\ee
(The predicted splittings of the latter do not appear in Table \ref{tabl} because they are presented in lattice units.)

From the Belle data, one obtains the ratio
\be
\frac{\hyperfine{2}}{\hyperfine{1}}=0.42\pm 0.09,
\label{belleratio}
\ee
in agreement with all of the above. On the other hand the ratio of Dobbs \etal.\
\be
\frac{\hyperfine{2}}{\hyperfine{1}}=0.73\pm 0.09
\label{dobbsratio}
\ee
is consistent only with Gray \etal.\ \cite{Gray:2005ur}.

\begin{figure*}
\includegraphics{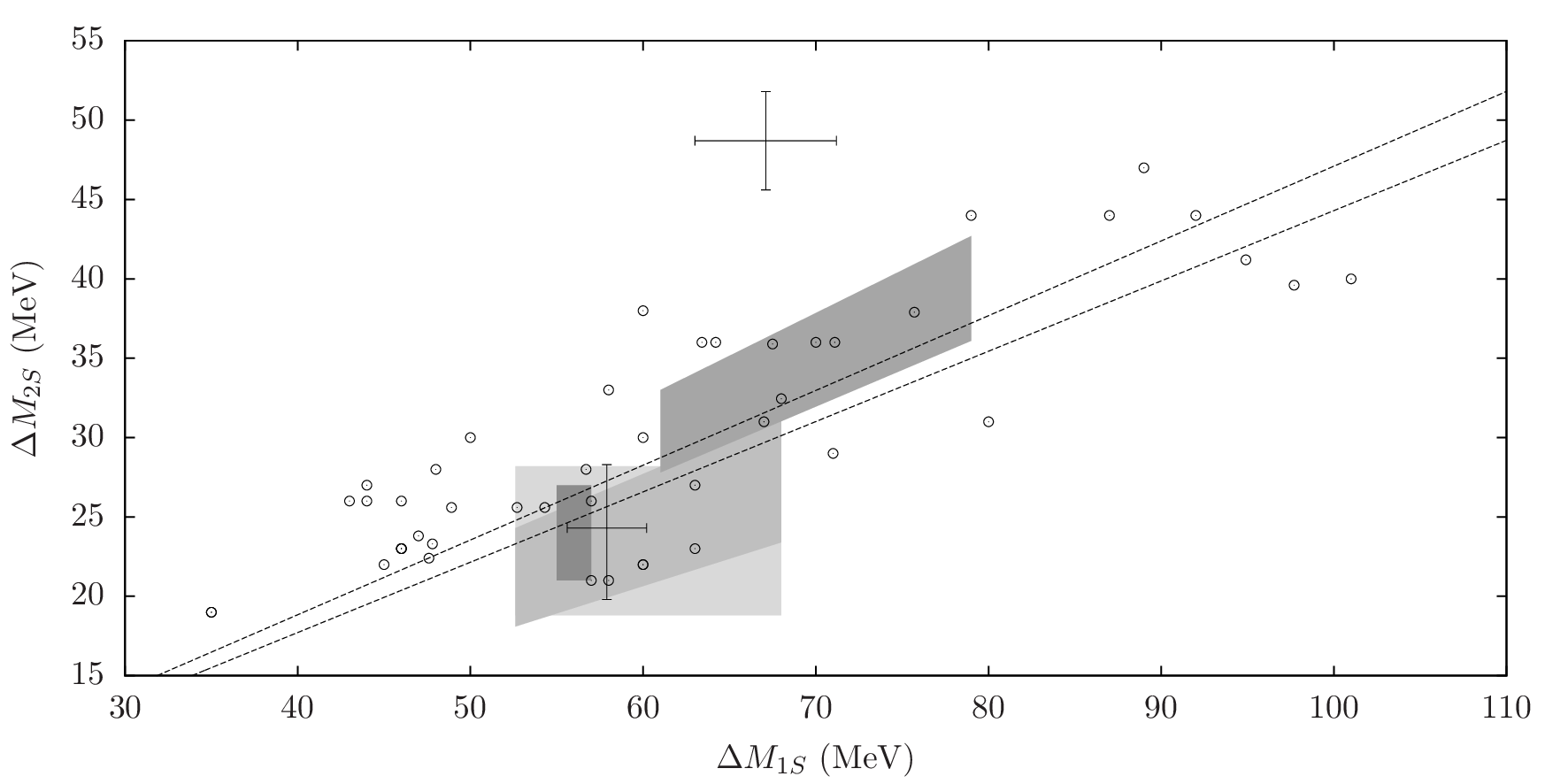} % for the PRD version, which requires a ps file containing everything
 \caption{Experimental data and theoretical predictions for the hyperfine splittings of 1S and 2S bottomonia. The data points of Dobbs \etal.\ and Belle are shown by the error bars. The predictions of lattice QCD  (Section \ref{lattice}) are shown by the shaded areas, the model-independent mass relation (Section \ref{relation}) by the broken lines, and potential models (Section \ref{potential}) by circles. }
\label{fig}
\end{figure*}

Figure \ref{fig} shows a plot of the 1S and 2S splittings of various theoretical approaches compared with experimental data. The measured splittings of Dobbs \etal.\ and Belle are shown by error bars, and the lattice predictions shown by the shaded areas. (The lines and circles show the predictions of the model-independent relation and potential models, discussed in the subsequent sections.) The skewed shapes are the results of Meinel and Dowdall \etal.\, using their measured 1S splittings and the measured ratios $\hyperfine{2}/\hyperfine{1}$. The remaining two data points are the other result of Meinel (which is normalised to the 1P tensor splitting) and that of Lewis and Woloshyn. For the sake of clarity, the results of Refs  \cite{Gray:2005ur,Burch:2007fj} with larger errors are not plotted.

\section{Model-independent relation}\label{relation}

In this section the experimental 2S splittings are compared to the predictions of a mass relation which is common to the simplest nonrelativistic potential models. The hyperfine splitting is controlled by the spin-spin term in the potential, which in tree-level perturbative QCD is proportional to a delta function in the quark separation,
\be
V_{SS}(\ve r)= \frac{32\pi\alpha_s}{9m^2}\delta^3(\ve r)\ve{S}_{q}\cdot \ve{S}_{\bar q},
\label{spinspin}
\ee
where $m$ is the quark mass and $\alpha_s$ controls the strength of the interaction. In perturbation theory the corresponding hyperfine splitting $\hyperfine{n}=M_{n\an3S1}-M_{n\an1S0}$ is proportional to the square of the radial wave function $R_{n\uS}$ at the origin:
\be
\hyperfine{n}=\frac{8\alpha_s}{9m^2}|R_{n\uS}(0)|^2.\label{qqmsplitting}
\ee
The $\epem$ width of an $n\an 3S1$ meson is also proportional to the square of the wave function at the origin, according to the Van Royen-Weisskopf formula \cite{VanRoyen:1967nq},
\be
\epemwidth{n}=\frac{4 {e_q}^2 \alpha^2}{M_{n}^2} |R_{n\uS}(0)|^2\left(1-\frac{16\alpha_s}{3\pi}\right),\label{qqmepemwidth}
\ee
where $e_q$ is the electric charge of the quarks in units of the fundamental charge, and the $\alpha_s$ term is the first QCD correction \cite{Barbieri:1975ki}. For $M_n$ one normally uses twice the quark mass, or the mass of the $n\an 3S1$ state; as discussed by Voloshin \cite{Voloshin:2007dx} the two prescriptions are equally valid in the nonrelativistic limit.

The model-independent wave function at the origin cancels in the ratio of hyperfine splitting to $\epem$ width:
\be
\frac{\hyperfine{n}}{\epemwidth{n}}=\frac{8}{9 e_q^2 \alpha^2}\left(\frac{M_n}{2m}\right)^2\frac{\alpha_s}{1-{16\alpha_s}/{3\pi}}.\label{ratio}
\ee
With the prescription $M_n=2m$, the experimental data on the 1S levels of charmonium and bottomonium satisfy the above with reasonable values of $\alpha_s= 0.29$ and $\alpha_s=0.22$, respectively.
Taking ratios of the above equation for 1S and 2S levels yields a relation (which depends upon the prescription for $M_n$) among the hyperfine splittings and $\epem$ widths of the 1S and 2S levels. With $M_n=2m$, 
\be
\frac{\hyperfine{2}}{\hyperfine{1}}=\frac{\epemwidth{2}}{\epemwidth{1}},
\ee
and with $M_n=M_{n\an3S1}$,
\be
\frac{\hyperfine{2}}{\hyperfine{1}}=\frac{\epemwidth{2}}{\epemwidth{1}}\left(\frac{M_{2\an3S1}}{M_{1\an3S1}}\right)^2.
\ee

Before applying the relations to bottomonia, it is interesting to test their validity for charmonia. The difference between the two prescriptions is quite significant, with $(M_{\psi'}/M_{J/\psi})^2=1.416$. The experimental data \cite{pdg12} gives the ratios
\bea
\frac{\hyperfine{2}}{\hyperfine{1}}&=&0.407\pm0.015,\\
\frac{\Gamma_{\epem\to\psi'}}{\Gamma_{\epem\to J/\psi}}&=&0.423\pm 0.018,
\eea
which are consistent with the relation taking the prescription $M_n=2m$.

For bottomonia, the difference between the two prescriptions is less significant, because $(M_{\Upsilon'}/M_{\Upsilon})^2=1.123$. This is another manifestation of the familiar situation that nonrelativistic predictions are more powerful for bottomonia than charmonia. The ratio of $\epem$ widths \cite{pdg12} is
\bea
\frac{\Gamma_{\epem\to\Upsilon'}}{\Gamma_{\epem\to \Upsilon}}=0.457\pm0.014.
\eea
The ratio \rf{belleratio} of hyperfine splittings at Belle therefore satisfies the model-independent relation using either prescription for $M_n$. On the other hand the ratio  \rf{dobbsratio} of Dobbs \etal.\ does not satisfy the relation using either prescription. The prediction of the relation (with $M_n=2m$) is shown by the broken lines in Figure \ref{fig}.

Notice also that the ratio of splittings predicted by the model-independent relation is consistent with all of the ratios  \rf{lattice[1]}, \rf{lattice[2]}, \rf{lattice[3]} and \rf{lattice[4]} computed on the lattice. From Figure \ref{fig},  it is also clearly consistent with the other lattice computations.

%so that the ratio of the 2S and 1S splittings is given by
%\be
%\frac{\hyperfine{2}}{\hyperfine{1}}=\frac{|\psi_{2\uS}(0)|^2}{|\psi_{1\uS}(0)|^2}
%\label{massratio}
%\ee
%Although the potential \rf{spinspin} arises from tree-level perturbative QCD, equation \rf{massratio} remains valid even after including one-loop corrections, because the hyperfine splitting is still proportional to the square of the wave function at the origin \cite{Buchmuller:1981aj}. 
%(However, non-perturbative effects  \cite{Voloshin:2007dx}.)

\section{Potential Models and related}\label{potential}

In this section the observed splittings at of Dobbs \etal.\ and Belle are compared to the predictions of potential models and related approaches. Table \ref{tabl} collects a range of such predictions, including ordinary nonrelativistic quark potential models with various functional forms (Cornell, logarithmic, power law, Buchmuller and Tye, Richardson's, screened); scalar, vector and mixed confining potentials; relativistic, relativised and semi-relativistic models; perturbative QCD and renormalisation group-improved potentials; and one model in which the potential is extracted from lattice QCD. The physics of the various approaches will not be discussed in detail here: the aim of this section is to draw more general conclusions as to the feasibility of fitting the disparate experimental results to potential model predictions.

Some of the approaches quote several predictions for the 1S and 2S splittings, corresponding to different parameter sets. In such cases, if a particular parameter set leads to a worse fit to all of the experimental data (PDG, Dobbs \etal.\ and Belle) than another parameter set, it is not included in the table. Those parameter sets which cannot be discriminated in this way are all included in the table.

The data in Table \ref{tabl} is plotted in Figure \ref{fig}. There is considerable spread in the predictions of both the 1S and 2S splittings. There is one data point which is consistent with the 2S splitting of Dobbs \etal.\, but the corresponding prediction for the 1S splitting is much larger than any of the experimental values. These splittings (the third pair of values quoted in the table for Bali \etal.\ \cite{Bali:1997am}) are obtained by solving the Schrodinger equation with a potential extracted from quenched lattice QCD, and incorporating an estimate of the correction due to the quenched approximation. The same approach applied to charmonia predicts 1S and 2S hyperfine splittings which are much larger than experiment.

All of the other models predict the 2S splitting smaller than that measured by Dobbs \etal., in most cases quite considerably smaller. Those which lie closest to 2S splitting of Dobbs \etal.\ all predict the 1S splitting much larger than experiment.

On the other hand many of the predictions  \cite{Bali:1997am,Bander:1983ew,Bander:1984vr,Buchmuller:1980su,Chen:2000ej,Fulcher:1988ca,Fulcher:1991dm,Godfrey:1985xj,Gupta:1986xt,Gupta:1989jd,Gupta:1995ps,Ito:1990he,Lengyel:2000dk,Motyka:1997di,Moxhay:1983vu,Rosner:1983jc,Pantaleone:1985uf,Radford:2007vd,Recksiegel:2003fm,Zeng:1994vj,Zhang:1991et} are consistent with the 2S splitting of Belle, among which the 1S splittings of several \cite{Bander:1983ew,Bander:1984vr,Motyka:1997di,Moxhay:1983vu,Rosner:1983jc,Recksiegel:2003fm} are also consistent with Belle, two  \cite{Godfrey:1985xj,Ito:1990he} are consistent with Dobbs \etal.\ and PDG12*, and the rest 
\cite{Bali:1997am,Buchmuller:1980su,Chen:2000ej,Fulcher:1988ca,Fulcher:1991dm,Gupta:1986xt,Gupta:1989jd,Gupta:1995ps,Lengyel:2000dk,Pantaleone:1985uf,Radford:2007vd,Zeng:1994vj,Zhang:1991et} are smaller than any of the measured values. 

The data in Figure \ref{fig} shows a positive correlation between the 1S and 2S splittings, as expected since both are controlled by the same parameter ($\alpha_s$) in the potential. It is plausible that models for which the ratio (but not the magnitude) of 2S and 1S splittings is in agreement with experiment can be brought into agreement also in magnitude by suitably adjusting $\alpha_s$. Of course $\alpha_s$ appears in other terms in the potential, so whether or not such an adjustment improves or spoils the overall fit to the spectrum will vary from model to model. Nevertheless it is noteworthy that the ratio of 2S to 1S splittings of Dobbs \etal.\ is not consistent with any of the models considered, as can be seen in Figure \ref{fig}: all of the data points would lie below the line of minimal slope between the origin and the error bars of Dobbs \etal. On the other hand, the measured ratio of Belle is  consistent with many models 
\cite{Bander:1983ew,Ito:1990he,Bander:1984vr,Ono:1982ft,McClary:1983xw,Chen:2000ej,Li:2009nr,Godfrey:1985xj,Eichten:1980mw,Moxhay:1983vu,Rosner:1983jc,Grotch:1984gf,Shah:2012js,Fulcher:1990kx,Gupta:1995ps,Gupta:1989jd,Recksiegel:2003fm,Motyka:1997di,Buchmuller:1980su,Fulcher:1991dm,Pantaleone:1985uf,Ebert:2002pp,Giachetti:2012tw,Eichten:1994gt,Badalian:2010uj,Radford:2007vd,Badalian:2009cx}. 

Finally it is interesting to consider separately the predictions of those models which came before and after the discovery of the $\etab{1}$. All of the models which post-date the discovery of the $\etab{1}$ 
\cite{Li:2009nr,Shah:2012js,Badalian:2010uj,Badalian:2009cx,Radford:2009qi,Giachetti:2012tw} are fit to the earlier 1S splittings measured at BaBar \cite{Aubert:2008ba,Aubert:2009as} and CLEO \cite{Bonvicini:2009hs}, or to the PDG averages at the time, and do not take account of the more recent, smaller values of Belle \cite{Collaboration:2011ch,Mizuk:2012pb} and Dobbs \etal.\ \cite{Dobbs:2012zn} shown in Table \ref{tabl}. Each of these models also predicts a 2S splitting somewhat larger than that observed at Belle. Presumably an updated fit in these models, which takes account of the smaller 1S splittings of more recent experiments, will decrease the predicted 2S splittings, possibly into agreement with Belle.

Among the predictions for the 1S splitting from models which pre-date the discovery of the $\etab{1}$ there are, in decreasing order of splitting: several which are larger than any of the measured values
\cite{Lahde:1998ee,Ono:1982ft,Eichten:1994gt,Bali:1997am,Fulcher:1990kx,Eichten:1980mw,Chen:2000ej,McClary:1983xw};
one \cite{Grotch:1984gf} consistent with PDG12;
two \cite{Godfrey:1985xj,Ito:1990he} consistent with the smaller PDG12* and Dobbs \etal.\ values;
several \cite{Bander:1983ew,Bander:1984vr,Ebert:2002pp,Motyka:1997di,Moxhay:1983vu,Rosner:1983jc,Patel:2008na,Recksiegel:2003fm} consistent with the even smaller value of Belle;
and quite a few \cite{Gupta:1982kp,Pantaleone:1985uf,Fulcher:1988ca,Bali:1997am,Gupta:1986xt,Recksiegel:2003fm,Buchmuller:1980su,Fulcher:1991dm,Lengyel:2000dk,Radford:2007vd,Gupta:1989jd,Zhang:1991et,Zeng:1994vj,Gupta:1995ps,Chen:2000ej} which are smaller than any of the measured values. The implication is that models which are not constrained to fit the $\etab{1}$ mass typically predict a smaller 1S splitting, which is interesting in light of the smaller measured values at recent experiments.

\section{The unquenched quark model}\label{unquenched}

The predictions of the model-independent relation and quark potential models discussed in Sections \ref{relation} and \ref{potential} ignore the effect of quark-pair creation (vacuum polarisation) on meson masses. In this section possible modifications to these predictions are considered in the framework of the unquenched quark model, which takes account of pair creation.

In the strong coupling picture of a heavy quark $\QQ$ meson, the creation of a light $\qq$ pair is manifest as a coupling to a pair of heavy-light mesons $\Qq$ and $\qQ$. Above threshold, this coupling leads to strong decay. Below threshold, it leads to a virtual $\Qq~\qQ$ component in the wave function and shifts the physical meson mass with respect to its bare mass.

Most models for the coupling assume that the $\qq$ pair is created in spin-triplet: where models differ is in the treatment of the spatial degrees of freedom. In the context of decay this assumption first appeared due to the requirement that the created $\qq$ pair has the  $\an 3P0$ quantum numbers of the vacuum \cite{Micu:1968mk}, but it is also follows from invariance arguments \cite{Alcock:1983gb} and is a feature of the flux-tube model of strong-coupling QCD \cite{Isgur:1983wj,Isgur:1984bm,Kokoski:1985is}. The first lattice QCD calculations of hybrid meson decay \cite{McNeile:2006bz} are also consistent with the spin-triplet hypothesis \cite{Burns:2006wz}.

In Refs \cite{Burns07production,Burns:2007hk} a general expression is presented for the matrix element for the coupling $\QQ\leftrightarrow\Qq~\qQ$ in models which assume spin-triplet pair creation. Barnes and Swanson \cite{Barnes:2007xu} apply the expression in perturbation theory to compute the mass shift due to quark-pair creation. Their key result, which applies in a symmetry limit, is that the shift is independent of the spin and total angular momenta of the $\QQ$. Thus, for example, the masses of bare states $n\an 1S0$ and $n\an 3S1$ are shifted equally and the hyperfine splitting is not affected. %More generally it implies that so that the effect of pair creation can be absorbed into a redefinition of the constant term in the potential. 
%, so that vacuum polarisation effects can be effectively be absorbed into the parameters of the ordinary quark model.

In practice this ideal limit is not realised: the symmetry is spoiled by the different masses of the bare $\QQ$ states, and of the heavy-light states $\Qq$ and $\qQ$ to which they couple. Bare states $\QQ$ with different spin and total angular momenta are shifted by different amounts, leading to induced spin-dependent splittings. The question relevant to this paper is how these effects modify the predictions for 1S and 2S hyperfine splittings. Refs \cite{Burns:2011fu,Burns:2011jv} address a closely related question, of how quark-pair creation modifies the hyperfine splitting of P-wave mesons. It turns out that the same arguments carry over to S-wave hyperfine splitting, so the approach and its conclusions are briefly summarised here. 

In the  nonrelativistic limit of the quenched quark model (which ignores pair creation) the hyperfine splitting
\be
\frac{1}{9}\left(M_{\an 3P0}+3M_{\an 3P1}+5M_{\an 3P2}\right)-M_{\an 1P1}
\ee
is exactly zero, a prediction strikingly consistent with experiment for 1P charmonia \cite{Dobbs:2008ec} and for 1P and 2P bottomonia \cite{:2011zp,Adachi:2011ji}. The induced mass shifts of the unquenched quark model might be expected to spoil the cancellation in the above linear combination, but remarkably this turns out not to be the case.

In Refs \cite{Burns:2011fu,Burns:2011jv} it is shown that there is a mechanism in place which keeps the P-wave hyperfine splitting small despite large corrections to the meson masses. The proof considers the effect of coupling to combinations of heavy-light pseudoscalar and vector mesons $\an1S0$ and $\an3S1$ in perturbation theory. (The shifts due to coupling to orbitally excited mesons are smaller, being suppressed by an energy denominator, and are in any case closer to the ideal limit of Barnes and Swanson  \cite{Barnes:2007xu} and so can largely be absorbed into a redefinition of model parameters.) Assuming only the creation of a $\qq$ in spin-triplet, the mass shift is expressed in terms of a power-series expansion in a parameter which is small provided that the mass differences among the bare and heavy-light mesons are small compared to the binding energy.

The result is that to first order in the expansion parameter, the effect of quark-pair creation is to suppress the hyperfine splitting with respect to that of the bare states by the factor $P_{\QQ}$, the probability that the physical meson is in the bare state $\QQ$ rather than the meson-meson state $\Qq~\qQ$:
\be
P_{\QQ}=1-P_{\Qq~\qQ}.
\ee
(Strictly, this is the probability computed using spin-averaged masses, though in practice it is similar to that using physical meson masses.)  Consequently, if the bare states have zero hyperfine splitting as in the nonrelativistic quark model, then so do the physical states, apart from small higher order corrections. In models in which the bare states have a non-zero hyperfine splitting, such as those which include relativistic corrections,  the effect of quark-pair creation is to reduce the splitting.

For the hyperfine splitting of S-wave mesons, it turns out that the same result applies. The hyperfine splitting $\hyperfine{n}'$ after including the effects of pair creation is suppressed compared to is bare value $\hyperfine{n}$ according to
\be
\hyperfine{n}'=P_{\QQ}\hyperfine{n} \label{prob}.
\ee
The derivation, which is analogous to but more straightforward than in the P-wave case, will be discussed in a future paper investigating generic features of the unquenched quark model. For the purposes of the present paper, its validity is verified by testing it against several model calculations in the literature. Three examples are shown in Table \ref{probtab}.

\begin{table}
\begin{ruledtabular}
 \begin{tabular}{llrrrr}
					&	& \multicolumn{1}{c}{$\hyperfine{n}$} & \multicolumn{1}{c}{$P_{\QQ}$} & \multicolumn{1}{l}{$\hyperfine{n}'$} &\multicolumn{1}{l}{($\hyperfine{n}'$)}\\ 
\hline
Barnes 					&1S & 118 & 0.69 & 84 & (81) \\ 
and Swanson \cite{Barnes:2007xu}	&2S & 49 & 0.51 & 25 & (25) \\ 
\hline
Kalashnikova \cite{Kalashnikova:2005ui}& 1S & 129 & 0.899 & 117 & (116) \\ 
&2S & 64 & 0.743 & 48 & (48) \\ 
\hline
Eichten \etal.\ \cite{Eichten:2004uh}& 1S & 120.7 & 0.966 & 117.1 & (116.6) \\ 
&2S & 67.2 & 0.702 & 46.3 & (47.1) \\ 
\end{tabular}
\end{ruledtabular}
\caption{Hyperfine splittings $\hyperfine{n}$ and $\hyperfine{n}'$ of charmonia in quenched and unquenched quark models respectively. The final column $(\hyperfine{n}')$ is the approximate splitting obtained from equation \rf{prob}. The probabilities $P_{\QQ}$ in equation \rf{prob} should strictly be computed using spin-averaged meson masses, whereas those which appear in this table are the quoted probabilities for the $n\an 3S1$ states computed using physical masses.}
\label{probtab}\end{table}

The first obvious consequence of equation \rf{prob} is that switching on quark-pair creation effects decreases the hyperfine splitting, since $P_{\QQ}<1$. This has immediate implications for the interpretation of the experimental data of Dobbs \etal.\ and Belle. 

In Figure \ref{fig}, all of the predictions for the 2S splitting in quenched quark model (with one exception) are smaller than that observed by Dobbs \etal. Incorporating quark pair creation will suppress these predicted splittings further, exacerbating the disagreement.

On the other hand, all of the predicted 2S splittings (with one exception) are consistent with or larger than that observed at Belle. In these cases, switching on pair creation has the potential to improve the overall agreement with Belle.

Equation \rf{prob} also has implications for the ratio of 2S to 1S splittings. Intuitively, one expects the probability $P_{\QQ}$ to be smaller for 2S  than for 1S states, since the former are closer to  threshold and therefore couple more strongly to heavy-light meson pairs. The examples in Table \ref{probtab} are all consistent with this expectation. Equation \rf{prob} then implies that switching on pair creation decreases the ratio of 2S to 1S splittings,
\be
\frac{\hyperfine{2}'}{\hyperfine{1}'}<\frac{\hyperfine{2}}{\hyperfine{1}}.
\ee
As well as the examples in Table \ref{probtab}, the inequality is satisfied by other model calculations \cite{Li:2009ad,Yang:2010am} which (because the probabilities $P_{\QQ}$ are not quoted) have not been included in the table. (In the case of ref \cite{Yang:2010am} the inequality is very close to an equality.) In the context of charmonia, Refs \cite{Eichten:2005ga,Eichten:2004uh} note that the 2S splitting is substantially reduced by pair-creation effects (compared to the 1S splitting), and that this improves the agreement of their model with experimental data; this observation is consistent with the above. In an earlier paper, Martin and Richard \cite{Martin:1982nw} point out that the 2S hyperfine splitting decreases because the $\psi(2\uS)$ couples to the lightest threshold $D\bar D$, whereas the $\eta_c(2\uS)$ does not.

Again, the effect makes it more difficult to reconcile the observations of Dobbs \etal.\ with theory. The ratios of 2S to 1S splittings of all of the quenched quark model predictions collected in Table \ref{tabl} and Figure \ref{fig} exceed the observed ratio of Dobbs \etal., as discussed earlier. Switching on the effects of pair creation will decrease the ratio further, worsening the disagreement. On the other hand, all of the ratios are consistent with, or larger than, the observations of Belle, so quark-pair creation could potentially improve the agreement.

Finally, the effect of switching on pair creation on the model-independent relation of Section \ref{relation} will be considered. As discussed, the hyperfine splitting $\hyperfine{n}$ of a given level is suppressed with respect to the quenched quark model value by a factor $P_{\QQ}$. For the $\epem$ width, the virtual photon couples to the $\QQ$ --- as opposed to $\Qq~\qQ$ --- component of the meson wave function. Ignoring the mixing between different $n$ levels induced by pair creation, which is typically small in models (see for example Refs \cite{Eichten:2004uh,Ono:1983rd}), the $\epem$ width of the unquenched quark model is related to that of the quenched quark model by
\be
\epemwidth{n}'=P_{\QQ}\epemwidth{n}.
\ee
The ratio \rf{ratio} of hyperfine splitting to $\epem$ width remains as it is in the quenched quark model
\be
\frac{\hyperfine{n}'}{\epemwidth{n}'}=\frac{\hyperfine{n}}{\epemwidth{n}}
\ee
and the model-independent relation holds.

\section{Conclusion}

On the basis of their predictions for the hyperfine splitting, each of the theoretical approaches discussed in this paper supports the interpretation of the Belle candidate as the $\etab{2}$. Among the lattice QCD results, there is only one quoted value which disagrees with the Belle hyperfine splitting, and this is shown to be consistent when normalised to more recent, smaller experimental values of the 1S splitting. In contrast, there is only one lattice result (with large errors) which is consistent with Dobbs \etal., and this is also consistent with Belle.

A model-independent relation, which is satisfied for charmonia and is expected to be more reliable for bottomonia, is in agreement with Belle but not with Dobbs \etal. The relation is shown to be consistent with lattice results.

Many potential models are consistent with Belle, and several others may plausibly be brought into agreement by adjusting $\alpha_s$ or taking account of mass shifts due to quark-pair creation. On the other hand, models consistently predict the hyperfine splitting much smaller than that of the Dobbs \etal.\ candidate, and it appears unlikely that adjusting model parameters or incorporating quark pair creation could bring them into agreement.

%\section{Appendix}

\bibliography{tjb.bib}                             

\end{document}